# Multiferroic Properties of Nanocrystalline BaTiO$_3$

R. V. K. Mangalam[†], Nirat Ray[‡], Umesh V. Waghmare[‡],

A. Sundaresan[†]* and C. N. R. Rao[†]

[†]*Chemistry and Physics of Materials Unit and DST unit on nanoscience,*

*Jawaharlal Nehru Centre for Advanced Scientific Research,*

*Jakkur P. O., Bangalore 560 064, India*

[‡]*Theoretical Sciences Unit,*

*Jawaharlal Nehru Centre for Advanced Scientific Research,*

*Jakkur P.O., Bangalore 560 064, India*

Some of the Multiferroics [1] form a rare class of materials that exhibit magnetoelectric coupling arising from the coexistence of ferromagnetism and ferroelectricity, with potential for many technological applications.[2,3] Over the last decade, an active research on multiferroics has resulted in the identification of a few routes that lead to multiferroicity in bulk materials.[4-6] While ferroelectricity in a classic ferroelectric such as BaTiO$_3$ is expected to diminish with the reducing particle size,[7,8] ferromagnetism cannot occur in its bulk form.[9] Here, we use a combination of experiment and first-principles simulations to demonstrate that multiferroic nature emerges in intermediate size nanocrystalline BaTiO$_3$, ferromagnetism arising from the oxygen vacancies at the surface and ferroelectricity from the core. A strong coupling between a surface polar phonon and spin is shown to result in a magnetocapacitance effect observed at room temperature, which can open up possibilities of new electro-magneto-mechanical devices at the nano-scale.

Ferroelectic materials exhibit spontaneous electric polarization that can be switched to its symmetry equivalent states with applied electric field. Ferroelectric $ABO_3$ perovskites are not only of fundamental interest but also of technological importance due to the strong coupling of their polarization with electric and stress fields. Switchability of the direction of polarization with applied electric field makes them useful in making non-volatile memories while their sensitivity to stress and strain fields is responsible for their use in micro-electromechanical systems. It will greatly widen the range of their applications, if the polarization is coupled with magnetic field as well. Such magnetoelectric coupling is a fundamentally interesting property allowed by the symmetry of multiferroic materials that exhibit ferroelectricity and ferromagnetism under the same physical conditions.

Ferroelectricity in $BaTiO_3$ (BTO) arises from the off-centering of the Ti ions with respect to a centrosymmetric cubic perovskite crystal. The fact that off-centering of the *B*-cation (eg. $Ti^{4+}$) originates from its $d^0$ electronic state,[9] contraindicates the possibility of magnetism that arises from local magnetic moments associated with the occupation of the *d*-states at the *B*-site. This renders the mechanism of multiferroicity vitally interesting and has attracted a considerable attention in recent years [4-6,10,11]. In perovskite materials such as $BiMnO_3$[12,13] and $BiFeO_3$[11,14] ferroelectricity arises from the stereochemically active 6*s* lone-pair of $Bi^{3+}$ ions, and the magnetism originates from interacting spins of the *d*-electrons of $Mn^{3+}$ or $Fe^{3+}$ ions. In manganites, such as $YMnO_3$, ferroelectricity is related to the geometry of its structure. While a superposition of two distinct types of charge ordering was shown to yield broken centrosymmetry and ferroelectricity in CMR materials,[5,15] spin cycloids[16] were found to be the source of polarization in TbMnO$_3$.[6,10]

It is known that ferroelectricity is suppressed at the nano-scale due to depolarization fields arising from the bound charges at the surface. For example, ferroelectricity in BTO nanoparticles disappears below a critical size (40 nm).[17,18] On the other hand, recent work has shown that ferromagnetism occurs in nanoparticles of the otherwise nonmagnetic oxides,[19] but decreases with increasing particle size. Magnetism in these nanoparticles, considered to arise from vacancies at the surface, is suggested to be a universal phenomenon. A recent finding of ferroelectricity in much smaller (12 nm) nanoparticles of BTO[20] motivated us to explore the simultaneous occurrence of ferromagnetism and ferroelectricity in BTO nanocrystalline sample where the former is expected to arise from the surface and the latter from the core. Here, we report the observation of room temperature ferromagnetism as well as ferroelectricity in the nano-crystalline BTO and uncover their origin using first-principles calculations.

Nanocrystalline BTO sample was prepared by polymer precursor method using BaCO$_3$ and titanium (IV) isopropoxide as starting materials.[21,22] Stoichiometric quantity of titanium (IV) isopropoxide was added to a mixture of ethanol and acetic acid (3:1 volume ratio) and stirred continuously. After two hours, barium citrate solution, prepared with required amount of citric acid and BaCO$_3$, was added. Polyvinyl alcohol solution was finally added to the mixture and followed by continuous stirring until the formation of a white sol. The sol was centrifuged and dried at room temperature. Thermogravimetric analysis of the as-prepared sample showed that the decomposition of organic components occurs around 500 ºC. In order to completely remove the organic part, the sample was heated at 700 ºC in oxygen. For characterization of ferroelectricity through measurement of P-E hysteresis loop, a dense nanocrystalline BTO sample was prepared by pressing the (40 nm) particles into pellets and heating at 1000 ºC for 1 hr. This step yields a dense nanocrystalline BTO that facilitates the measurement of ferroelectric hysteresis loop.

FESEM image of samples heated at 700ºC and 1000 ºC show rods containing assemblies of nanoparticles with an average diameter - 40 nm (Fig. 1a) and a dense nanocrystalline BTO sample (Fig. 1b) with an average diameter of about 300 nm respectively. The FESEM image of the bulk BTO sample (Fig. 1c), obtained by sintering the pressed nanoparticles at 1200 ºC, shows the grain size to be around ~ 2$\mu$m. Room-temperature X-ray diffraction pattern of all the three samples showed the tetragonal (P4*mm*) structure, with the lattice parameters, $a$ = 4.0047(2) Å, & $c$ = 4.0226(4) Å, $a$ = 3.9968(1) Å, & $c$ = 4.0298(1) Å and $a$ = 3.9951(1) Å, & $c$ = 4.0377(1) Å, respectively. The increase in $c/a$ ratio with particle size is consistent with increasing tetragonality and magnitude of off-center distortion of Ti-ions from their cubic symmetry.[23]

Measurement of magnetization in the BTO particles heated at 700ºC and 1000ºC clearly show ferromagnetism at room temperature (see Fig. 2a & 2b). The coercivity increases from 95 Oe to 435 Oe with increase in particle size from 40 to 300 nm, which may be due to change in the distribution of defects at the surface and shape anisotropy. On the other hand, the saturation magnetization reduces from 0.0025 to 0.0012 emu/gm with increasing particle size and finally the bulk BTO sample failed to exhibit magnetic hysteresis as expected of a bulk BTO. It exhibits diamagnetic behavior at room temperature as shown in Fig 2b. Since there are no magnetic elements involved in the preparation of BTO nanoparticles and there was no contamination of magnetic materials during the process of magnetization measurements, the ferromagnetism is intrinsic to the nanocrystalline BTO. The observed ferromagnetism is consistent with the suggestion that nanoparticles of the otherwise non-magnetic oxides are ferromagnetic.[19] The origin of ferromagnetism may lie in the magnetic moments arising from the oxygen vacancies on the surfaces of the nanoparticles as suggested earlier.[19] It is, therefore, understandable that an increase in the particle size caused by sintering eliminates the magnetism due to the decrease in the surface to volume ratio. In fact, it has been shown theoretically that point defects such as cation or anion vacancies in insulators can create magnetic moments.[24,25] Further, it was shown that the range of exchange interaction between the magnetic moments and the critical concentration of vacancy determine the magnetic ground state of a material, for example $HfO_2$, where neutral Hf vacancies induce magnetic moments on neighboring oxygens.[26]

The polarization (P-E) hysteresis loop measured for the 40 nm sample is (Fig. 2c) similar to that reported for 50-100 nm BTO particles.[27] This kind of P-E loop does not quite represent the true ferroelectric nature of the particles, as one can expect such behavior in the P-E measurement carried out on cold pressed nanoparticles. It requires a local probe such as Electric Force Microscope (EFM) to see switching of polarization in individual nanoparticles. In fact, with the use of EFM technique it has already been shown that nanoparticles of BaTiO$_3$ with an average size of as small as 12 nm is ferroelectric.[20] However, a clear ferroelectric nature of polarization hysteresis is seen (Fig. 2d) in the sintered particles (300 nm). For a drive voltage of 2 kV, the hysteresis loop recorded at a frequency of 100 Hz shows a remnant polarization ($P_r$) value of 2.04 $\mu C/cm^2$, a maximum polarization ($P_{Max}$) of 8.42 $\mu C/cm^2$ and a coercive field ($E_c$) of 10 kV/cm. For the bulk BTO, these parameters are; $P_r$ = 14.5 $\mu C/cm^2$, $P_{Max}$ = 23.4 $\mu C/cm^2$ and $E_c$ = 22 kV/cm.

The temperature dependence of the dielectric constant (see Fig.3a) of 300 nm nanocrystalline BTO sample at various frequencies exhibits dielectric anomalies around 230 K and 300 K, which correlate respectively with the ferroelectric phase transitions of BTO from the low-temperature rhombohedral to the orthorhombic structure followed by the orthorhombic to tetragonal phases. The coupling between magnetization and electric polarization in the BTO nanocrystalline sample is manifested in the magnetocapacitance (MC): $\Delta\varepsilon(H)/\varepsilon(0) = [\varepsilon(H)-\varepsilon(0)]/\varepsilon(0)$. We observe a positive MC of fairly large magnitude (10 %) near room temperature in an applied magnetic field of 1 T (see Fig. 3b for its temperature dependence in the interval 77-360 K), which remains positive down to 77 K. The increase of MC above room temperature should be associated with the tetragonal to cubic transition that is expected to be around 400 K.

We now develop an understanding of these observations using first-principles density functional theory (DFT) calculations, which have proven to be quite effective in the determination of the origin of ferroelectricity in $BaTiO_3$,[28] focusing here on identification of the origin of its ferromagnetism at nano-scale. Simplifying the geometry of nanocrystalline BTO and their structure, we consider here a slab of $BaTiO_3$ of 1.2 nm thickness, consisting of seven [100] atomic planes and terminated with $TiO_2$ planes. We use a periodic supercell that consists of the slab and vacuum layers with in-plane periodicity of 2×2 unit cells of $BaTiO_3$. In the absence of any defects, such a slab is a good insulator as each plane is charge neutral (either BaO or $TiO_2$) and does not exhibit any magnetism as all the $Ti^{4+}$ cations are in the $d^0$ state.

Our first-principles calculations are based on a spin-dependent density functional theory calculations using a standard plane-wave code Quantum ESPRESSO[29] with a local density approximation[30] to interaction energy of electrons. We use ultra-soft pseudo-potentials[31] to represent the interaction between ions and electrons, and include semi-core *s* and *p* states of Ti and Ba explicitly in the valence. An energy cutoff of 30 Ry (180 Ry) on the plane wave basis was used in representation of Kohn-Sham wave-functions (density). Most calculations involved supercells with 70 atoms; phonons were determined for a smaller supercell with 24 atoms using a frozen phonon method. The supercell with 70 atoms consists of 2 × 2 unit cells in the plane of the surface and allows for reconstruction, particularly in the presence of oxygen vacancies at the surface. Brillouin zone integrations were sampled with a Monkhorst-Pack mesh[32] of *k*-points that is equivalent to a 4×4×4 mesh for the primitive cell of $BaTiO_3$. Structural optimization was carried out with BFGS algorithm to minimize energy using Hellman-Feynman forces.

From the earlier experimental[19] and theoretical[25] work, it is known that magnetism in oxide nanoparticles arises from vacancies. Oxygen vacancies are not uncommon in oxides of the type BTO. To assess the site preference of oxygen vacancies, we determined fully relaxed spin-dependent ground state structures for two configurations, one with oxygen vacancies in the bulk, and another with oxygen vacancies at the surface of the slab. With the choice of the supercell we have used, the in-plane concentration of vacancies is 12.5 atomic % ( < 5 atomic % in cell). We find that the configuration with oxygen vacancies at the surface is lower in energy by about 1.2 eV per oxygen vacancy. The magnitude of this energy difference is much larger than any errors in the DFT calculations and implies a much greater abundance of oxygen vacancies at the surface than in the bulk at room temperature ($\Delta E/k_B T = 48$; note that the entropy of vacancy sites in the bulk is higher due to larger number of configurations available, but its contribution to free energy diminishes at nano-scale).

We now assess the stability of ferromagnetic and antiferromagnetic states in the presence of oxygen vacancies both in the bulk and at surfaces. As the coordination of oxygen is two (both being Ti atoms), its vacancy is expected to result in occupation of d-states of primarily the neighbouring Ti atoms with an electron each. FM and AFM ordering have been simulated by initializing spins on these two Ti ions to parallel and anti-parallel configurations respectively. In either case, we find the ferromagnetic state to be lower in energy by more than 10 meV with 2 $\mu_B$ magnetic moment per oxygen vacancy. The origin of this can be understood from interpretation of the density of states (see Fig. 4) with exchange interactions. Upon introduction of oxygen vacancies, states with Ti $d$- character at the bottom of the conduction band get populated with a total of two electrons per oxygen vacancy. These states are relatively extended (more than one nm from the surface, see Fig. 5a, b) and their energies split up due to spin-dependent interaction. As these states are quite extended in the plane of the surface, they allow mediation the magnetic interactions necessary for a long-range order.[25,26] The ferromagnetic superexchange involves two different *d*- states (*xz* and *yz*) populated with electrons of parallel spins and Hund's coupling favors it to antiferromagnetic superexchange involving the same *d*- state populated with electrons of anti-parallel spins (see Fig 4). Consistent with its greater stability, the density of states in the FM configuration exhibits a dip or a pseudo-gap at the Fermi level.

To estimate the effects of oxygen vacancies on ferroelectricity, we have determined the energetics of structural distortions corresponding to local energy minima in which all the Ti atoms in the supercell are off-centered along a given direction (say, (001), (110) and (100)). In the ferromagnetic state of the nano-slab with oxygen vacancies at the surface, we find the ferroelectric phase polarized along (110) axis to be the lowest in energy, followed by the ones polarized along (100) and (001) directions respectively. The energy lowering with ferroelectric distortions along each of the three directions is noticeably lower for the slab with reflection symmetry in the horizontal plane ($\sigma_z$, oxygen vacancies at both surfaces), than that for the asymmetric slab with oxygen vacancy at only one of the two surfaces. While the lowering of energy with ferroelectric distortions in slabs is about three times weaker than in the bulk, ferroelectricity does seem to survive in the nano-thin slabs of $BaTiO_3$ with oxygen vacancies at the surface. Ferroelectric structural distortions mainly involve off-centering of Ti and O atoms and they are much smaller at the surface than in the interior of the slab (see Fig. 5c).

To assess the local stability of the completely relaxed FM and FE structure, we have determined Γ point phonons (note that they include phonons corresponding to Γ, X, M and R modes of the bulk). We find that all the phonons are stable in the FM state, with four of them below 100 cm$^{-1}$, which is expected of a ferroelectric material.[33] The frequency of most of these phonons changes by at most 0.5 % when the magnetic ordering changes from FM to AFM type, indicating a rather weak spin-phonon coupling. However, we find an unstable phonon ($\omega = 170\,i$ cm$^{-1}$) in the AFM state that has a strong overlap (40 %) with a phonon at 300 cm$^{-1}$ in the FM state, indicating its very strong coupling with spin. This mode (see Fig. 5d) is confined to the surface of the slab (like the magnetization density of the AFM state) and has some components of the other low-energy phonons of the FM state as well. Its coupling with spin should have therefore observable consequences when these soft modes show a strong temperature dependence, ie. near the ferroelectric transitions. In particular, changes in magnetization with applied magnetic field would result in shifts in the frequency of these soft polar modes and hence in its contribution to dielectric response. This surface mode (Fig. 5d) is thus responsible for the magneto-capacitive anomaly observed here experimentally.

In conclusion, the present study demonstrates a new kind of multiferroicity in nanocrystalline ferroelectric oxide. The multiferroic nature is rendered possible by the surface magnetism of the nanocrystalline BTO. Interestingly, the ferroelectric and magnetic properties are coupled as shown by the observation of magnetocapacitance. It is interesting to ponder on the possible existence of both ferroelectricity and ferromagnetism occurring at the surfaces of nanoparticles of the otherwise non-ferroic oxides.

**Figure captions**

**Figure 1.** FESEM image of BTO sample heated to (a) 700°C (40 nm) (b) 1000 °C (300 nm) and (c) bulk BTO sample obtained by sintering the nanoparticles at 1200 °C.

**Figure 2.** Magnetic hysteresis curves recorded at room temperature for BTO particles with an average size of (a) 40 nm (b) 300 nm, confirming that these particles are ferromagnetic. It is clearly seen from (b) that the bulk BTO is diamagnetic as expected. P-E hysteresis measured at room temperature for (c) 40 nm and (d) 300 nm particles.

**Figure 3.** (a) Temperature dependent dielectric constant and (b) magnetocapacitance of 300 nm nanocrystalline BTO.

**Figure 4.** Mechanism of magnetism in $BaTiO_3$ with oxygen vacancies. Top two panels show the density of electronic states of AFM and FM states of the configuration with oxygen vacancies at the surfaces. In the bottom half, a sketch of superexchange interactions shows that the same *d*-state is involved in the AFM state, while two different *d*-states are involved in the FM state. The latter is more stable than the former due to the Hund's coupling that favors parallel alignment of spins of electrons in different *d*-states.

**Figure 5.** Magnetization density and low energy structural distortions of nano-slabs of $BaTiO_3$. Barium, Titanium and Oxygen atoms are shown as blue, yellow and red spheres respectively, and the slab extends periodically to infinity in the plane perpendicular to z-

axis (vertical direction [001]). Isosurfaces of magnetization density at 10 % of its peak values (red and dark blue isosurfaces indicate positive and negative magnetization respectively) shown for AFM state (a) and FM state (b) of nano-slab with oxygen vacancy on the top [001] plane reveal that the magnetization penetrates about 1 nm from the surface. Cylindrical symmetry of magnetization density at Ti sites in the FM state (b) gives additional evidence for the two *d*-orbitals *xz* and *yz* involved in FM exchange. In (c), arrows indicate atomic displacements (within a factor) that link a ferroelectric FM state polarized along [001] direction with the reference centrosymmetric FM state of nano-slab with oxygen vacancies at both the surfaces. It is clear that polar off-centering atomic displacements are very small at the surfaces. In (d), the unstable phonon mode found in the AFM state of nano-slab with 5 atomic planes and oxygen vacancies on both the surfaces: it is localized at the surface and couples strongly with spin-ordering at the surface.

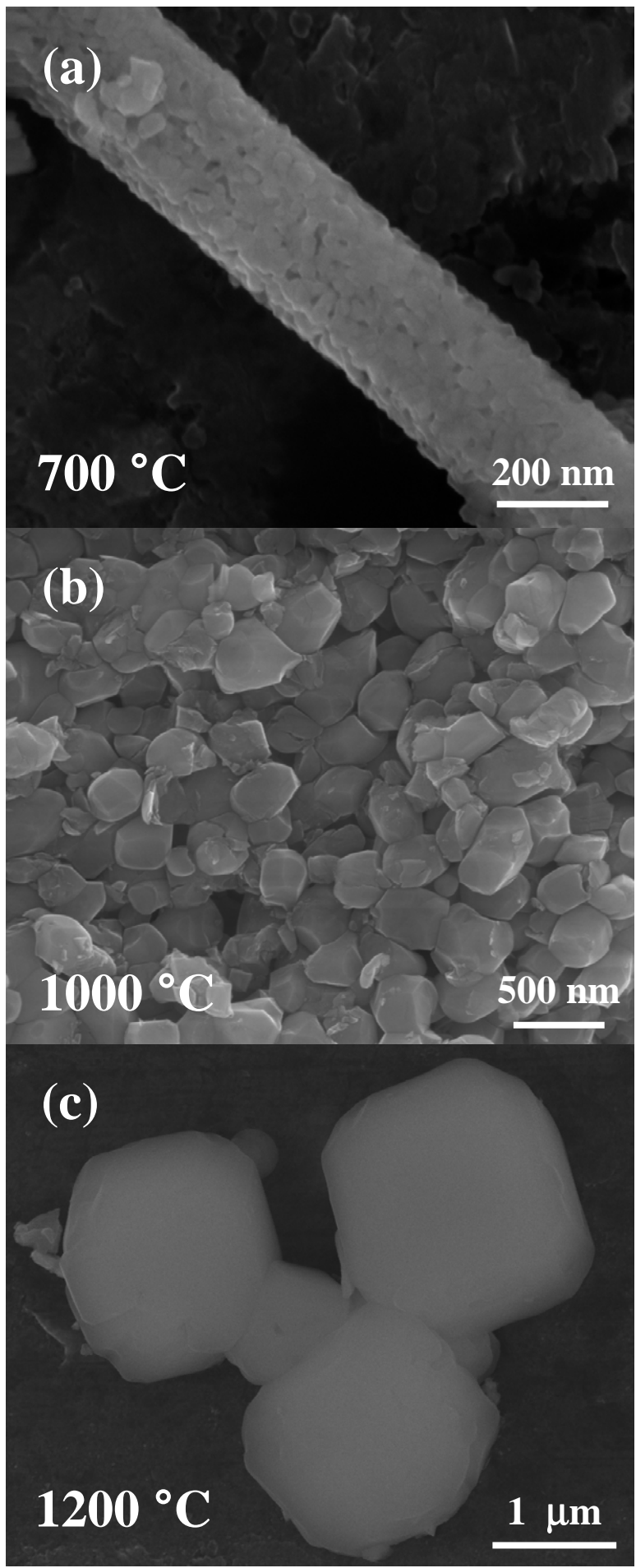

**Figure 1**

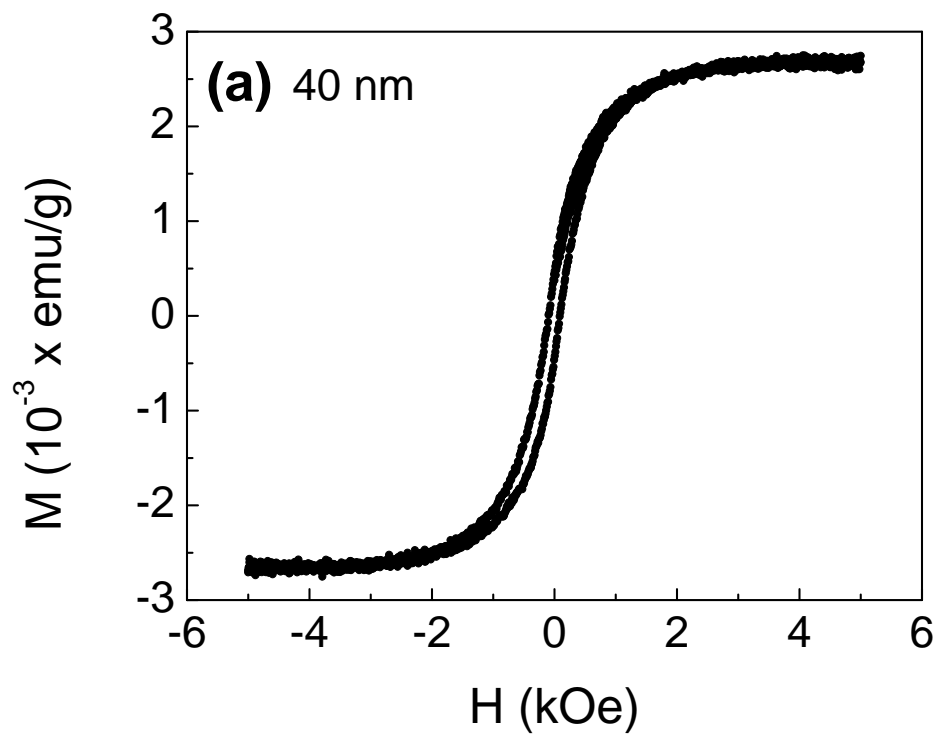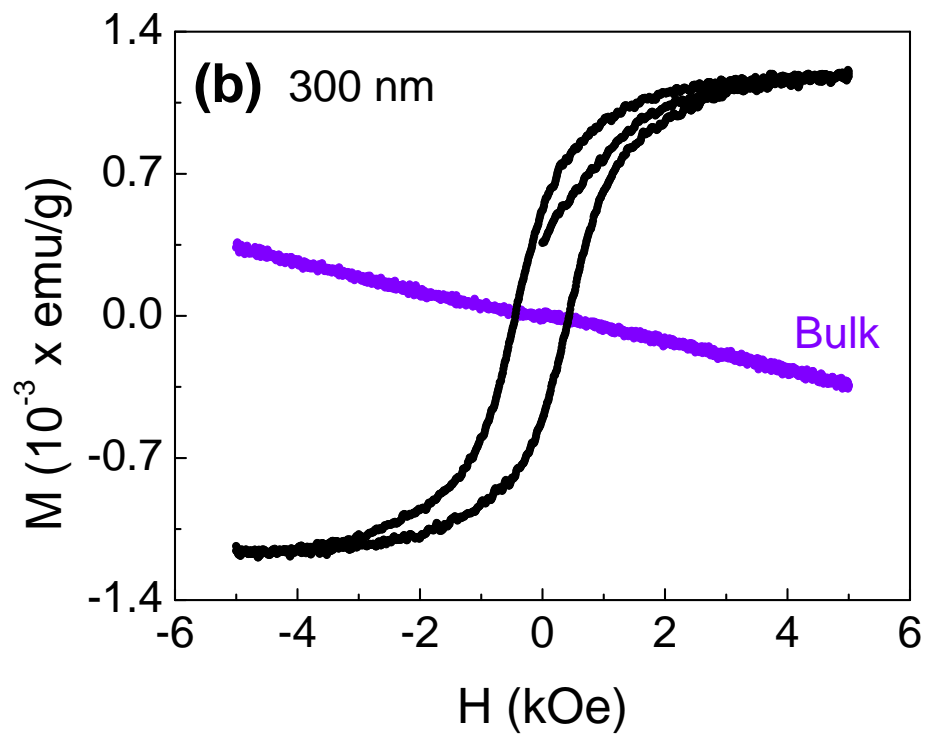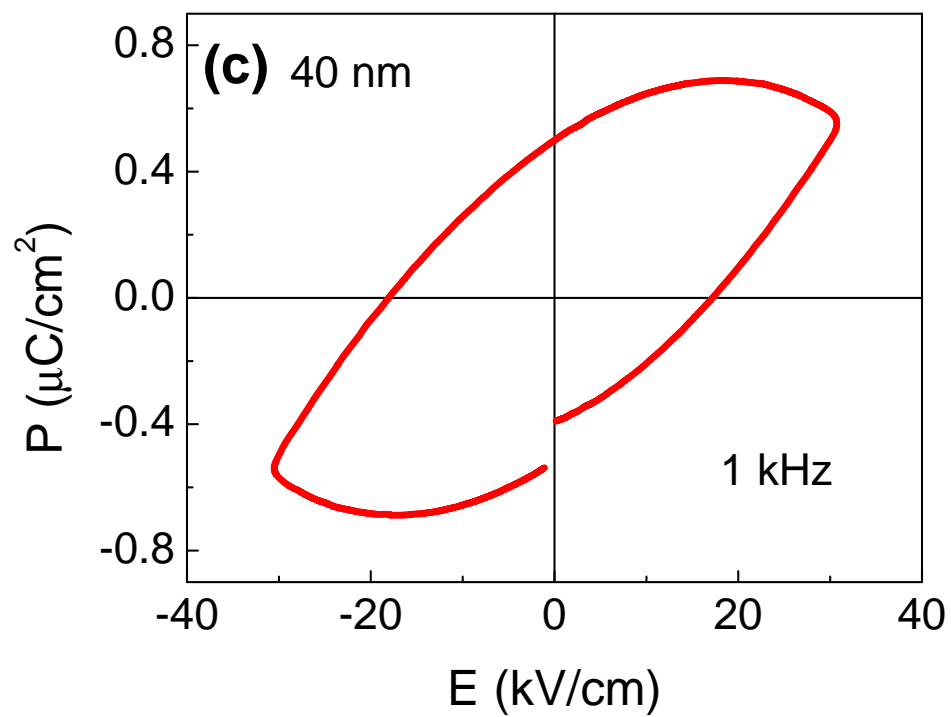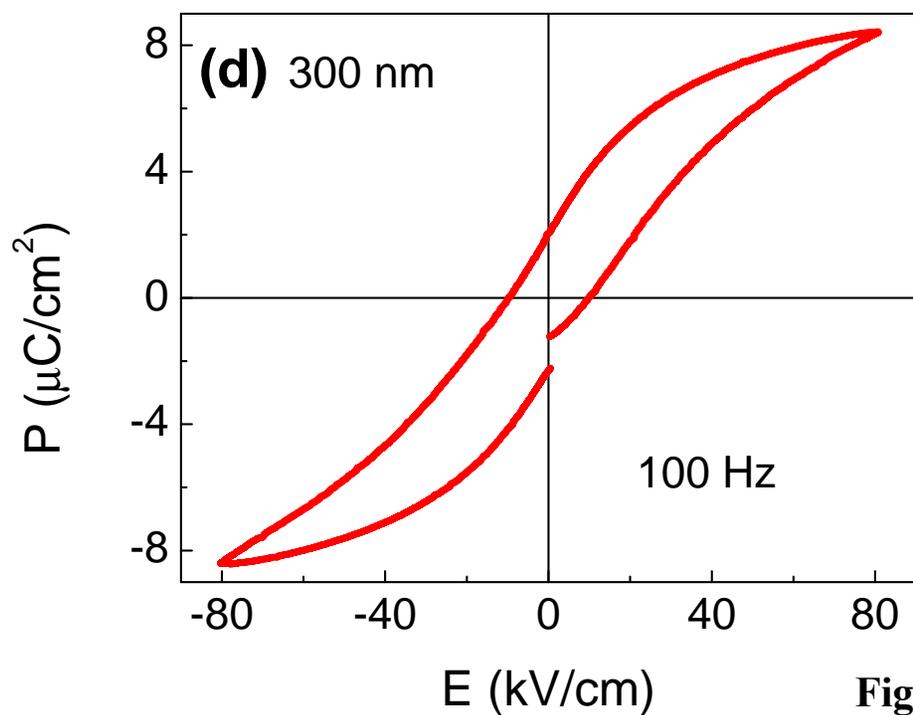

**Figure 2**

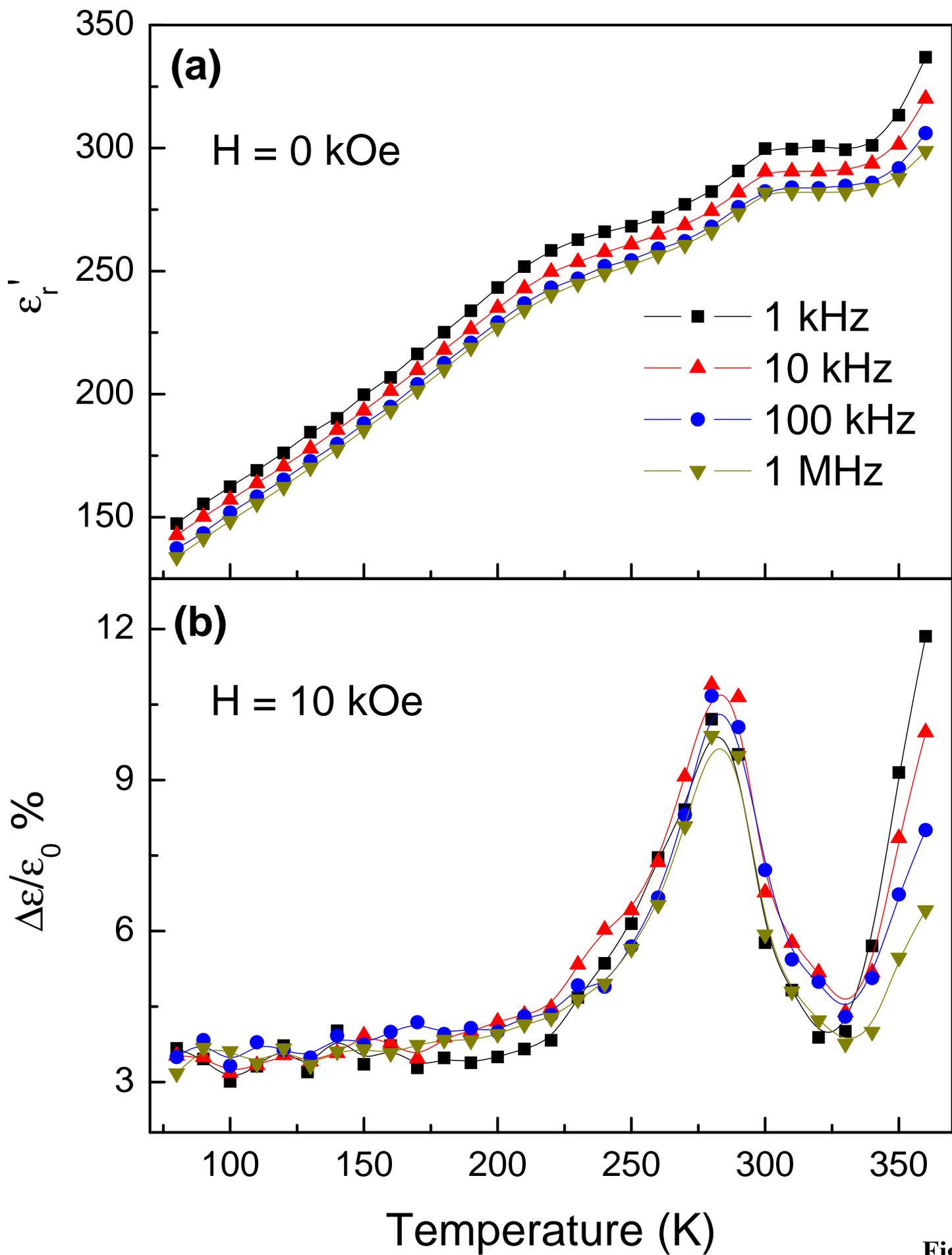

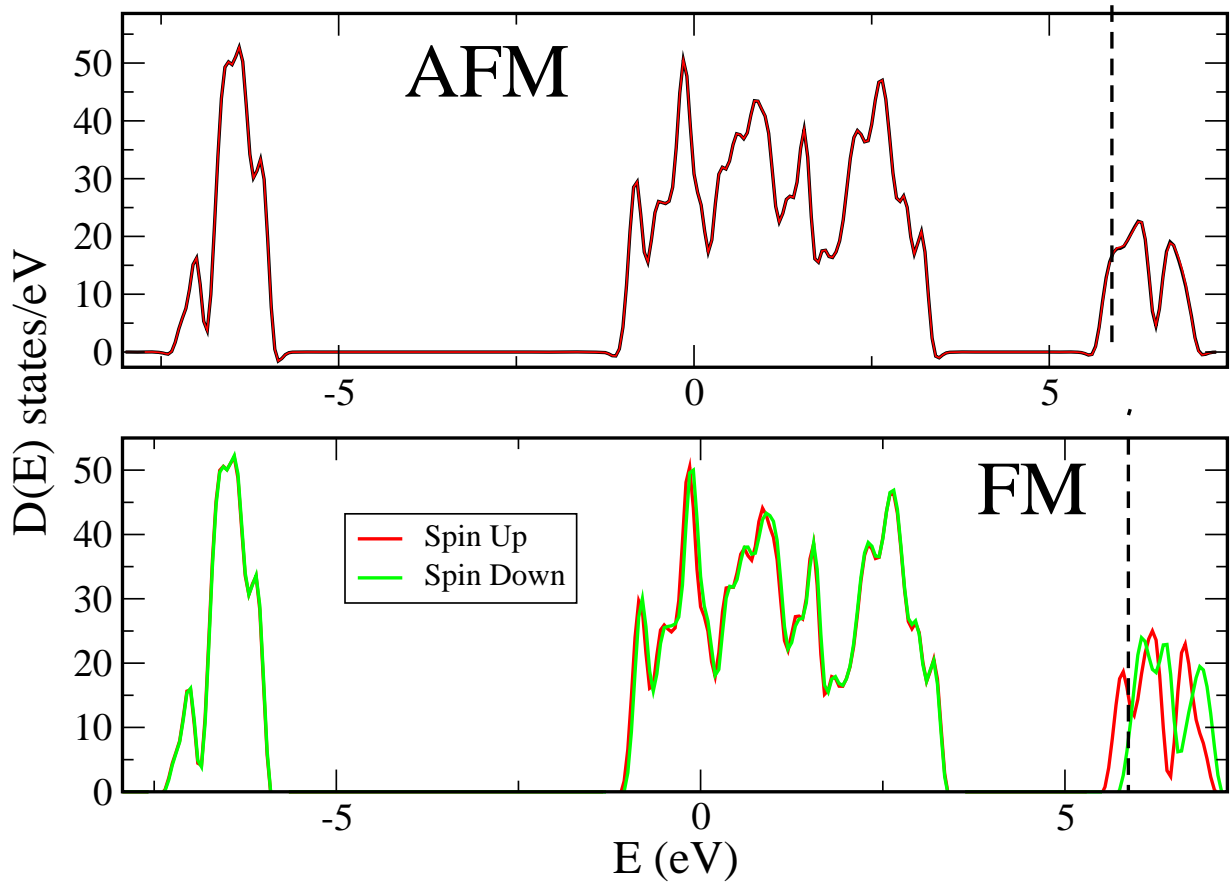
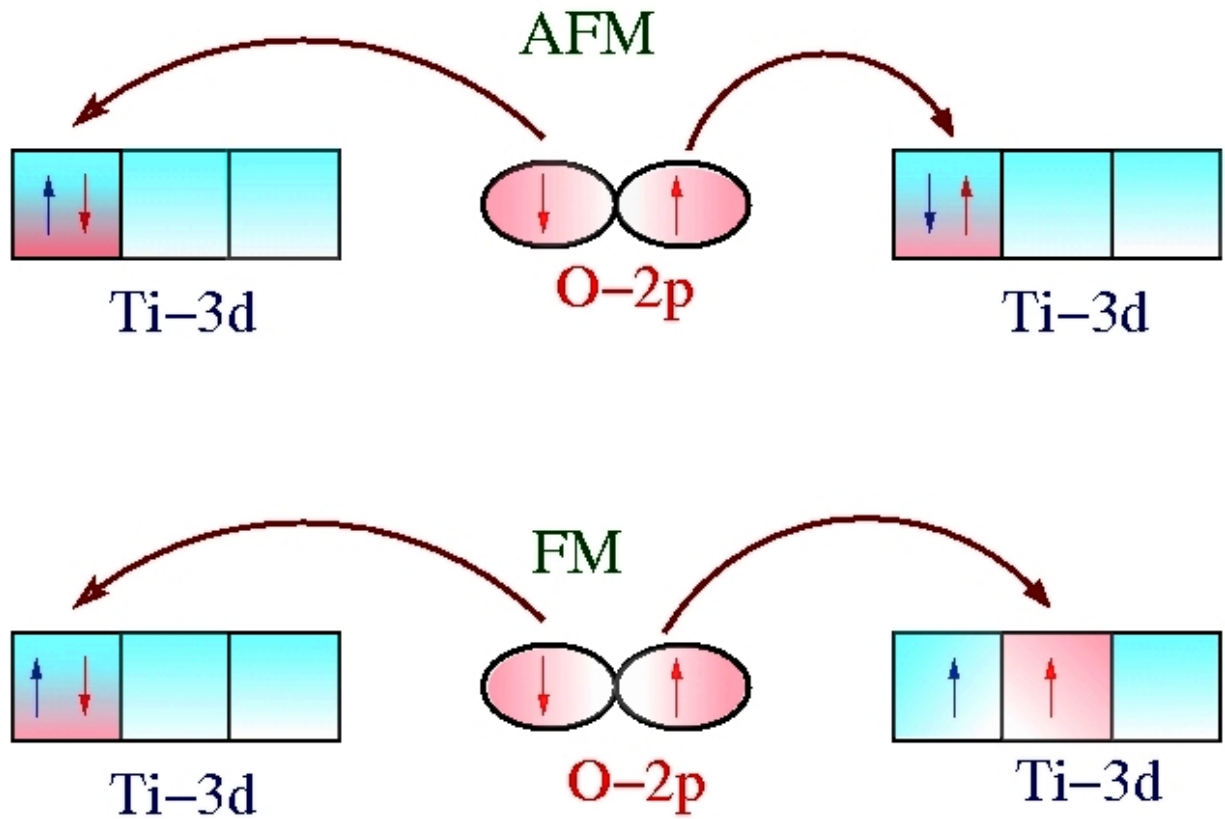

**Figure 4**

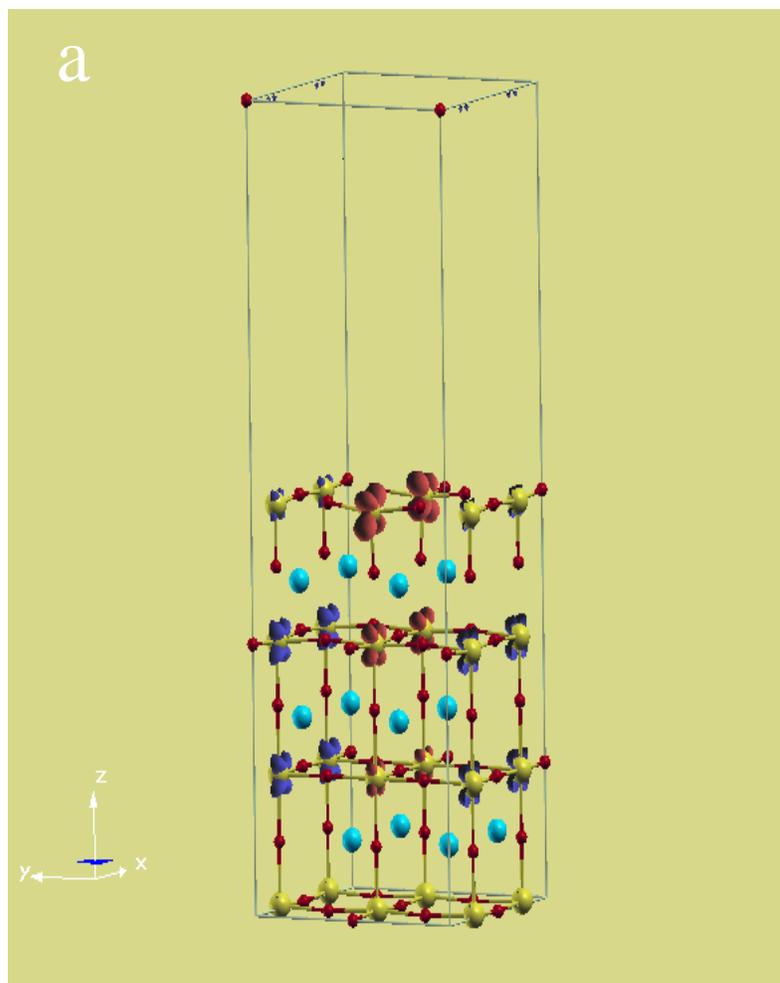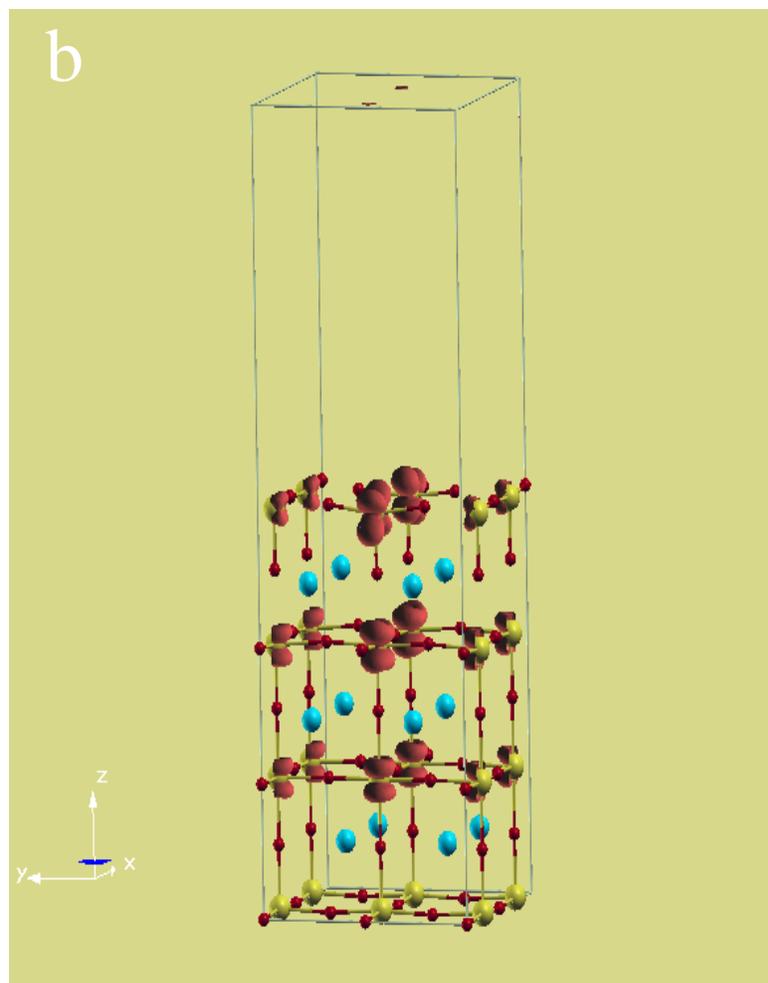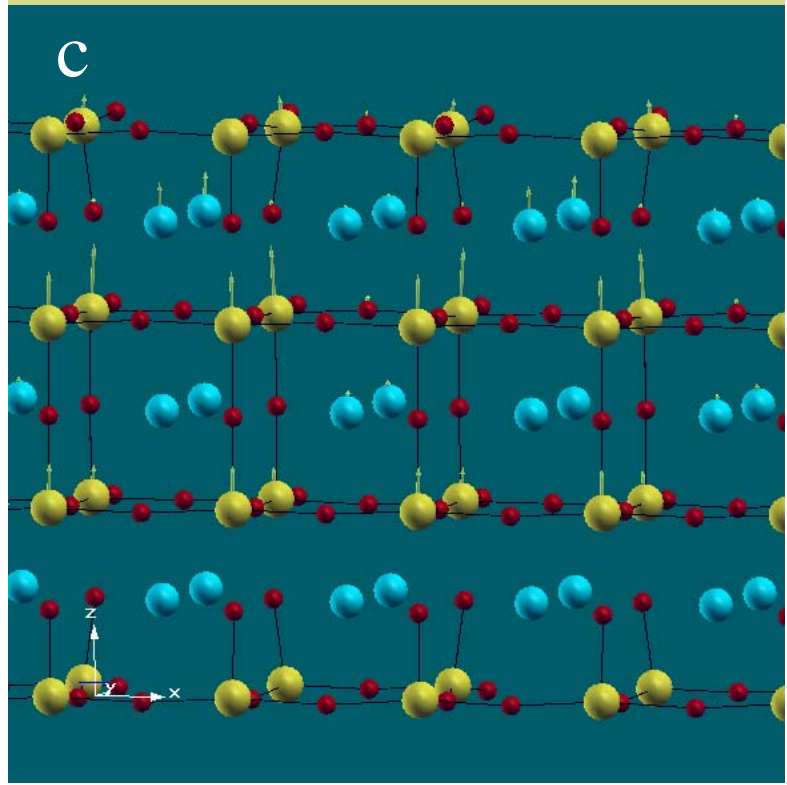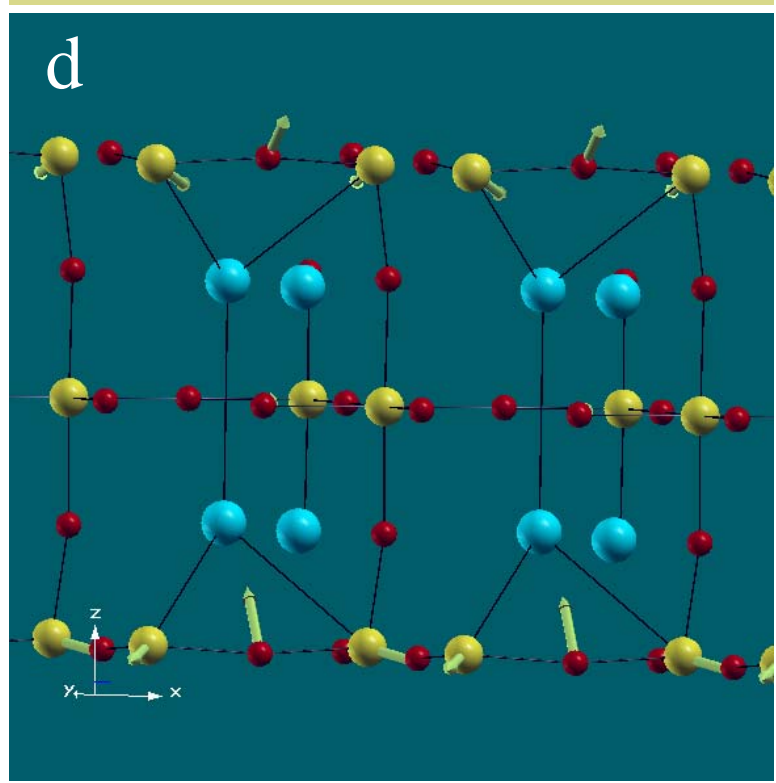

**Figure 5**